\documentclass[11pt]{article}
\usepackage{hyperref}
\pdfoutput=1
\begin{document}

\title{Influence of a local change of depth on\\ the behavior of bouncing oil drops}%
\author{Remi Carmigniani, Simon Lapointe,\\ Sean Symon \& Beverley J. McKeon\\
\\ 
Graduate Aerospace Laboratories,\\California Institute of Technology,\\ Pasadena, CA 91125, USA}%
\maketitle
\begin{abstract}

The work of Couder \textit{et al} \cite{couder2005} (see also Bush \textit{et al} \cite{bush1, bush2}) inspired consideration of the impact of a submerged obstacle, providing a local change of depth, on the behavior of oil drops in the bouncing regime. In the linked videos, we recreate some of their results for a drop bouncing on a uniform depth bath of the same liquid undergoing vertical oscillations just below the conditions for Faraday instability, and show a range of new behaviors associated with change of depth.

This article accompanies a fluid dynamics video entered into the Gallery of Fluid Motion of the 66th Annual Meeting of the APS Division of Fluid Dynamics.
\end{abstract}

\section{Introduction}

An experiment was conducted to study the bouncing behavior of silicone oil drops on a vertically oscillating shallow tray of the same liquid. The simple bouncing, period-doubled bouncing and walking regimes as well as chaotic and intermittent behaviors were observed and occurred under conditions very similar to those of previous work~\cite{couder2005, jfm2006, bush1, bush2}.

The influence of depth on the trajectories of drops walking over a submerged obstacle was investigated. The local change of depth associated with the obstacle led to a range of possible trajectories, including straight crossing, reflection from the aft face of the obstacle and trapping of the droplet above the obstacle. A reduction in walking velocity occurred in all cases. The different behaviors were dependent on the approach velocity of the drops; the change of depth influenced the decay time of the standing waves generated during bouncing, thus producing changes to the walking speed and direction.

\section{Experimental Set-up}

A circular aluminum tray of 6 inches in diameter was filled with silicone oil (\emph{Clearco} PSF-50cSt) and mounted on a vibration exciter. The depth of the oil film was greater than 4 mm so that the Faraday waves were independent of this parameter. An accelerometer was installed on the side of the tray to measure the amplitude and frequency of motion.

The influence of a local change in depth was studied by gluing a submerged obstacle, namely an aluminum strip with height $l_w = 7$ mm, width $e=3.5$ mm and length 120 mm, to the bottom of the tray along its diameter. In order to ensure that the drops would walk toward the obstacle, two guides were placed forming a triangular shape with the obstacle~\cite{eddi2009}. The height of these guides exceeded the maximum depth of the oil so that drops would reflect off them if they got close. The tray was filled with silicone oil until the surface was 2 mm above the obstacle, $h_w = 2$ mm

To create drops of the desired size, a needle or a sharp toothpick ($\approx 2$ mm in diameter) was plunged inside the oil tray and rapidly pulled out. When the resultant liquid ligament detached from the surface, it either produced one small droplet or a set of droplets with distributed sizes via the Plateau-Rayleigh instability, depending on the rate at which the liquid bridge was elongated. This technique produced small drops (generally $0.5 < D < 1.1$ mm).

\section{Videos}

High and low resolution versions, respectively, of the video entries to the 2013 Gallery of Fluid Motion are provided in ancillary files\\ \\
\vspace{12pt}
\textit{V102356\_InfluenceLocalChangeDepth\_BouncingDrop} and\\
\vspace{6pt}
\textit{V102356\_InfluenceLocalChangeDepth\_BouncingDrop\_SMALL}. \\

In these movies, the following phenomena are shown:
\begin{itemize}
\item Simple Bouncing (bath of uniform depth)
\item Double Bouncing (bath of uniform depth)
\item Walking (bath of uniform depth)
\item Walking (local change in bath depth), with the following specific behaviors
\begin{enumerate}
\item Crossing
\item Rebound
\item Trapping
\end{enumerate}
\end{itemize}

The new behaviors associated with a change in bath depth are explained in the context of pilot waves, as described for the uniform depth case by Bush \textit{et al} \cite{bush1,bush2}.

\end{document}